# Architecture of a First-Generation Commercial Quantum Network

Duncan Earl, K Karunaratne, Jason Schaake, Ryan Strum, Patrick Swingle, Ryan Wilson

Qubitekk, Inc., Vista, CA, 92081 USA

*Abstract* – **We present the architecture and near-term use cases for a first-generation commercial quantum network. We define the foundational hardware and software elements required to operate and manage the network. Finally, we discuss the configuration of this network for near-term consumer applications and propose how the network can support the broader technical goals of the quantum information science community.**

## I. INTRODUCTION

Quantum networks are necessary to accelerate the development, adoption, and integration of quantum solutions. Experimental quantum networks are being developed throughout the world as scientists and engineers work to interconnect quantum computing [1–4], quantum communication [5–10], and quantum sensing [11–13] devices. Although quantum networks share many similarities with their classical counterparts, they also require unique and specialized hardware and network architecture considerations [14–16]. In this paper we describe the first quantum network developed and constructed from commercial quantum devices and intended for use by quantum product developers for commercial activities. This "commercial quantum network" supports multiple users and can be managed and maintained by classical network administrators. Capable of generating, distributing, and measuring photonic qubits among multiple users, the network exhibits a scalable architecture that can expand to include more users, more resources, and quickly incorporate newly commercialized quantum devices.

## II. DEFINITION OF A COMMERCIAL QUANTUM NETWORK

The high-level functionality of any quantum network is to coherently interconnect geographically distributed quantum devices. To carry quantum information from one network node to another, photonic qubits are often employed due to their compatibility with existing free space and fiber optic channels [17–20]. Much as a classical optical network will generate, distribute, and measure optical pulses to convey classical data (or bits), a quantum network must have the resources and architecture to generate, distribute, and measure photonic quantum states (or qubits) across a network.

To meet these high-level functional requirements, we submit that it is the responsibility of the quantum network to provide, at a minimum:

- A source of photonic qubits
- An optical layer for transmitting photonic qubits
- All-optical switches capable of routing photonic qubits
- Hardware for maintaining the coherency of distributed photonic qubits
- A device for measuring the state of a photonic qubit
- Electronics for precisely timing detected qubits correlated across the network
- A classical communication layer for equipment configuration and data sharing
- Software for monitoring and configuring the network

This specialized equipment - provided by the quantum network - is a "common use" resource that can be used by all users to transmit information and interact with the quantum network. This common-use equipment obviates the need for quantum product developers to recreate complete, end-to-end solutions. If the common-use equipment is reconfigurable, different configurations of the equipment can serve different solutions. The greater the quantity of the common-use resources, the greater the number of network users and unique solutions that can be supported.

A quantum network must consist of at least one optical channel for transmitting qubits between users. For a commercial quantum network, multiple parallel optical channels can be provided to create a "qubit bus." An N-wide qubit bus between network users supports solutions that require parallel qubit processing, such as distributed quantum processing [21, 22].

For a commercial quantum network, the common-use equipment, software, and network management must be achieved through the integration of commercially released products offered and supported by industry vendors. Only recently has this been possible due to the growth of the quantum component industry. Reliable commercial qubit sources, single photon detectors, quantum-compatible fiber optic switches, and a variety of other devices and software are now available for procurement and inclusion in a commercial quantum network [23].

Commercial quantum networks can be further defined by how they share user access to network resources. A commercial quantum network, owned and maintained by a commercial entity, can provide access to networked quantum resources through a subscriber-based model. Users, or subscribers, of a commercial quantum network are charged a fee for connection, access, and configuration of the network. Similar to the operation of early cellular telecommunication networks, users can be charged a flat-fee or a use-based fee depending on various revenue models. Recently, asset owners of traditional fiber optic networks have begun exploring how quantum networks can augment their operations and expand services to customers and communities. A commercial quantum network incorporates these business considerations into its operation and provides the necessary mechanisms to support revenue generation [24].

In this paper we define the architecture of a commercial quantum network that supports qubit communications across a 4-qubit wide fiber optic bus. The network architecture is scalable and permits the number of subscribers to be increased as network resources are increased. We designate this scalable, first-generation architecture as the Bohr-IV Quantum Network.

## III. NETWORK ELEMENTS AND SYMBOLS

To describe the architecture of the Bohr-IV Quantum Network, we must first describe the foundational elements of the network hardware. These network elements include:

- optical fibers and switches
- the quantum Equipment Hub
- the client Quantum Node
- and the Control Center

This section will describe these four foundational elements while the following section will define the interconnectivity between these elements.

### A. Optical Fibers and Switches

The Bohr-IV Quantum Network is a fiber optic network that requires twelve optical fibers to be routed between all network elements. These twelve fibers consist of a Primary fiber bundle (5 fibers), a Secondary fiber bundle (5 fibers), and a LAN bundle (2 fibers).

The primary and secondary bundles each consist of four optical fibers dedicated to carrying qubits, while a fifth fiber carries classical optical pulses used in the precision timing of events. Two fibers are used to establish a classical local area network (LAN) between all classical computing, memory, etc. devices on the network. The symbols for representing these fibers and their routing are shown in Figure 1.

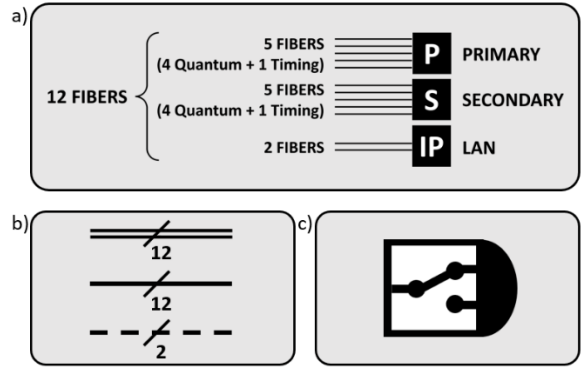

Fig. 1. **Optical fibers and switch elements:** (**a**) Breakout showing use of twelve optical fibers between network elements and showing symbols for Primary, Secondary and LAN connections. (**b**) Representations used for representing optical fiber bundles that contain multiple strands, including hub-to-hub (top), hub to node (middle), LAN only connections (bottom). (**c**) Symbol for an all-optical, quantum compatible fiber optic switch.

All-optical fiber optic switches are used to electronically configure the routing of the quantum and timing channels. The required switch configurations include: 60x60, 20x20, and 8x24. These switches are considered "quantum compatible" in that they do not incorporate any photo-to-electric conversion and they include features, such as ultralow channel crosstalk, enhanced suppression of internal photon scattering, polarization insensitive optical efficiency, length-matched optical fibers, and low-loss interconnects.

## B. Quantum Equipment Hubs

For maintenance and management purposes, it is convenient to group common-use quantum equipment into protected locations physically separated from the network users. Just as classical network hardware (e.g., switches, repeaters, etc.) are located in protected data rooms, a quantum network requires specialized quantum equipment to be located in protected quantum Equipment Hubs.

The symbol for a quantum Equipment Hub is shown in Figure 2.

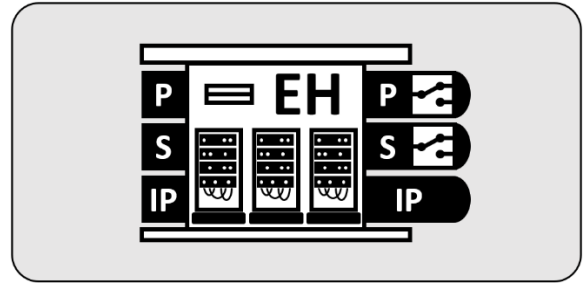

Fig. 2. **Quantum Equipment Hub symbol:** The quantum Equipment Hub contains input and output interconnects to the twelve optical fibers interconnecting network nodes. Within the Equipment Hub, common-use quantum devices are located and monitored within equipment racks.

The Equipment Hub has Primary, Secondary, and LAN interconnects that engage with the 12-strand optical fiber bundle described previously. Within each Equipment Hub, instrument racks located in environmentally controlled spaces are used to house the equipment. For the Bohr-IV Quantum Network, these racks include the following equipment:

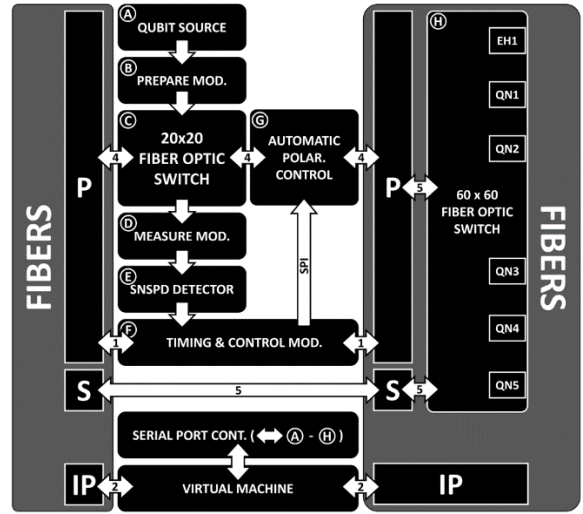

Fig. 3. **Components in quantum Equipment Hub:** Includes photonic qubit sources, Prepare Modules to prepare various types of qubits, a 20x20 fiber optic switch to configure the interconnection between equipment, Measure Modules to measure the quantum states of qubits, SNSPD detectors to detect single photon events, automatic polarization controllers to maintain polarization across optical fibers, a Timing & Control Module to correlate detected events and regulate polarization controllers, and a 60x60 fiber optic switch to control routing between the Equipment Hub and other network nodes.

All devices in the Equipment Hub are commercially available products with warrantied performance specifications for a defined operating environment. Adjustment of each device's operating settings (e.g., a switch's connectivity configuration) is made through serial commands passed to the device through a serial port controller and attached virtual machine. Although details on the principles of operation and performance requirements for each device are outside the scope of this architecture-centric paper, a high-level description of the major components is included.

The commercial qubit source consists of four continuous-wave bi-photon sources based on Type II spontaneous parametric down-conversion in a bulk periodically-poled KTP crystal. The temperature tuned bi-photon output produces photon pairs with orthogonal polarization that is degenerate at 1570nm and has a bandwidth of approximately 2nm.

The output of the qubit sources can be directed through one of three Prepare Modules via fiber optic switches (not shown in Figure 3) connected immediately before and after the three Prepare Modules. An example of a Prepare Module is a device containing non-polarizing beamsplitters that converts the bi-photon output from each source into post-selected, polarization entangled qubits for use on the network. Alternatively, a Prepare Module can contain polarizing beamsplitters that convert the bi-photon sources into heralded qubits. Generally, the Prepare Modules are intended to be passive optical devices that prepare the photonic qubits in various common ways.

Prepared qubits are connected to an any-to-any, 20x20 fiber optic MEMS switch. The switch is capable of routing qubits to devices internal to the Equipment Hub, or to optical fibers that lead to nodes upstream and downstream of the Equipment Hub. For qubits directed internally, one of three Measure Modules can be selected via fiber optic switches (not shown in Figure 3) immediately before and after the three Measure Modules. An example of a Measure Module is a polarizing beamsplitter that can measure the polarization of an incoming photon. It should be noted that, depending on the configuration of the 20x20 switch, the input to the Measure Modules can be qubits generated within the Equipment Hub or qubits generated from outside the Equipment Hub (i.e., in other network nodes). As with the Prepare Modules, the Measure Modules are generally intended to be passive optical devices that perform a measurement of the received qubit's quantum state.

The output of the Measure Modules is an 8-channel superconducting nanowire single photon detector (SNSPD) optimized for use at 1570nm. The commercial SNSPD detector produces electrical pulses for every photon detected on a given channel and these outputs are monitored by the Timing & Control Module to correlate detected events across the network. The commercial Timing & Control Module is an FPGA-based device that monitors detection events and uses eighteen SFP lasers to send multiplexed timing signals to other Timing & Control Modules on the network. A high-speed timing protocol allows correlations between events detected by other Timing & Control Modules to be measured and counted in real time.

For qubits routed external to the Equipment Hub, a 4-channel Automatic Polarization Control (APC) Module ensures that qubit polarization, in any basis, is preserved across four parallel fiber optic quantum channels. The APC uses commercial, multi-stage, lithium niobate modulators to actively correct for polarization rotations in the optical fiber and uses single photon correlations – detected at the Timing & Control Modules - to provide feedback for an APC control algorithm.

Finally, an any-to-any, 60x60 fiber optic switch is used to connect the Equipment Hub to other nodes on the network. The 60x60 switch is configurable, allowing network topologies to be adjusted for various applications.

## C. Client Quantum Nodes

Quantum Nodes allow users to integrate their equipment and implement solutions on the quantum network. Each Quantum Node is equipped with a Quantum Network Interface Console – or QNIC – that provides physical connectivity and the ability to reconfigure the network. The QNIC allows the user to physically access photonic qubits and real-time signal data through the Primary and Secondary fiber optic channels, while providing limited access to the network's LAN layer for transmitting network configuration requests and receiving real-time count data and network status updates.

The QNIC consists of a Timing & Control Module along with other devices to support a user's optical and electrical interconnects. Any user or third-party equipment connected to the QNIC must meet the Quantum Physical Layer Connectivity Requirements (QPHY) for the deployed network. The non-proprietary QPHY connectivity requirements define the optical, electrical, and data transmission requirements for interfacing with the QNIC. This includes data encoding protocols (e.g., 8b/10b), baud rates, timing protocols, electrical signal levels, system commands, etc. that must be observed by connected electronics, embedded software, and user applications. Additional requirements for specific network applications may be required for integrating application-specific hardware (e.g., QKD solutions) onto the network. The symbol for a client Quantum Node is shown in Figure 4.

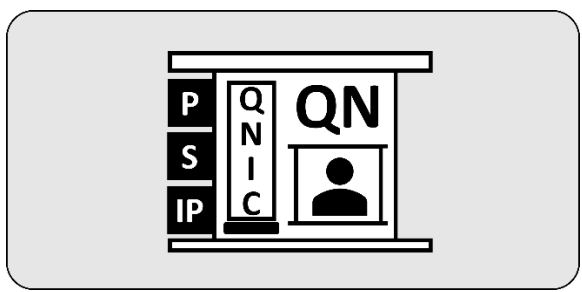

Fig. 4. **Client Quantum Node symbol:** The client Quantum Node identifies where users connect to and access the quantum network. A Quantum Network Interface Console (QNIC) is used by a user to physically connect to the quantum network.

## D. Control Center

Even a small quantum network can consist of dozens of common-use devices with hundreds of settings and parameters. A commercial quantum network must ensure that these devices are configured and continually monitored to ensure the network is performing as designed and that new network configurations are properly configured and stable. A Control Center is a manned facility where the network's equipment is monitored, new network configurations are instantiated, and network maintenance and billing activities are coordinated. The Control Center is the recipient of user requests and oversees the scheduling of network configurations. In addition, the Control Center is responsible for storing and authorizing access to data generated on the network. The symbol for the Control Center is shown in Figure 5.

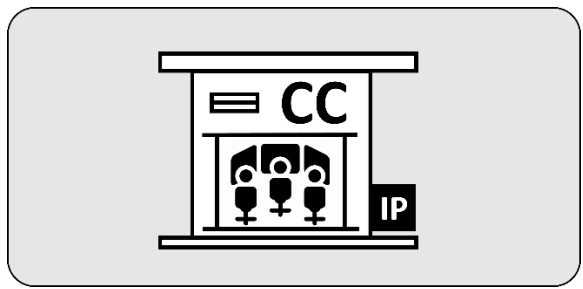

Fig. 5. **Control Center symbol:** The Control Center configures and monitors the commercial quantum network.

## IV. NETWORK ARCHITECTURE

With the key elements of a commercial quantum network identified, and symbols developed for each, the architecture of a first-generation commercial quantum network can be presented. Figure 6 illustrates the fiber optic connections between Equipment Hubs, client Quantum Nodes, and the Control Center. This represents the minimum scale implementation of the Bohr-IV Quantum Network. As shown in Figure 6, twelve optical fibers connect Equipment Hubs in a ring topology, while five client Quantum Nodes – also connected through twelve optical fibers – connect to each Equipment Hub in a hub-and-spoke topology.

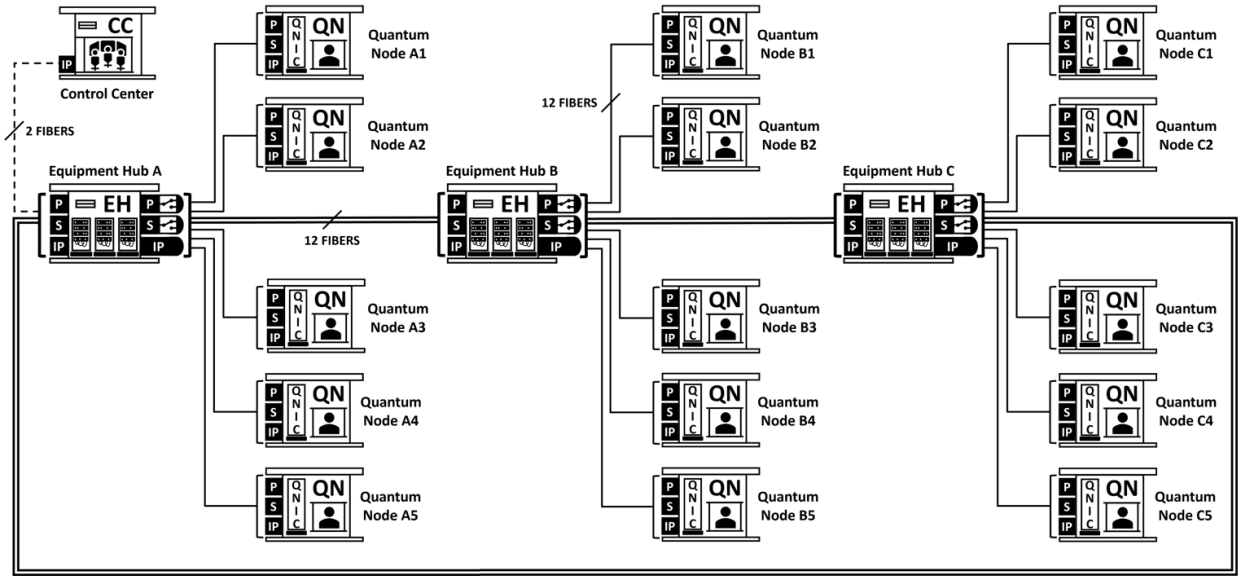

Fig. 6. **Bohr-IV Quantum Network Architecture:** Equipment Hubs are connected in a ring topology with five client Quantum Nodes connected to every Equipment Hub in a hub-and-spoke topology. This hybrid ring/spoke architecture is scaled by adding additional Equipment Hubs, which in turn adds additional Quantum Nodes.

The network architecture for the Bohr-IV Quantum Network allows the number of clients to be scaled by installing additional Equipment Hubs. Each new Equipment Hub adds up to five more clients to the network while also increasing the common-use resources available to all users.

Users design network configurations optimized for their applications using commercial software. Network configurations designed and optimized for a particular application are compiled into a network configuration request. Compiled network configuration requests are submitted by the user to the Control Center for implementation during a scheduled time. The Control Center monitors all equipment and configures all devices to implement an approved network configuration during a scheduled implementation window.

Configurations of the network can interconnect all users of the network or only a small subset. Likewise, a single user can utilize all available common-use resources (in all Equipment Hubs) or those resources can be distributed across multiple users. The reconfigurable topology is made possible by the 60x60 fiber optic switch located in each Equipment Hub. As shown previously in Figure 3, the 60x60 switch connects the internal resources of an Equipment Hub to five spoked Quantum Nodes (QN1-QN5). In addition, this switch connects the Equipment Hub to the next Equipment Hub in the ring topology (EH1). By adjusting the configuration of the switch, Quantum Nodes can be added or dropped from the final network topology. To support this functionality, the commercial 60x60 switch is constructed as shown in Figure 7.

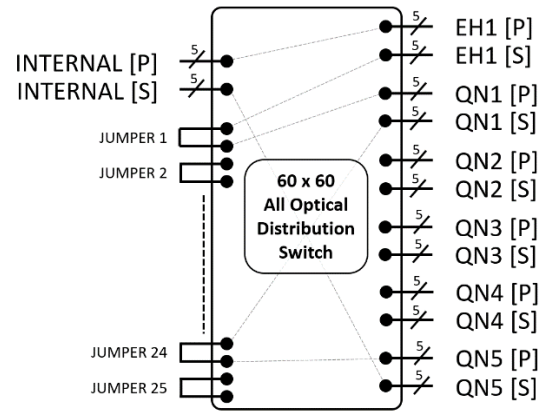

Fig. 7. **Construction of 60x60 Optical Switch:** Configuration of the optical switch can drop or add network users. Jumpers on the left side of the switch are used to interconnect Quantum Nodes on the right. Dotted lines provide an example of left to right connections.

As can be seen in Figure 7, "jumper" connections on the left side of the switch are used to interconnect (i.e., drop or add) the fibers on the right side associated with each Quantum Node.

The Bohr-IV Quantum Network architecture provides topological flexibility while also providing users with access to common-use, commercial quantum resources. This approach captures one of the primary goals of the first-generation commercial quantum network, which is to establish the critical infrastructure for quantum solutions and eliminate the need for developers to provide all aspects of an end-to-end quantum solution.

V. CONFIGURATION AND MANAGEMENT

Two high-level software packages are required to configure and operate the Bohr-IV Quantum Network. These include:

- User Design & Interface software
- Control Plane software

The User Design & Interface software is a graphical interface used by a user to design, simulate, and compile a desired network configuration. This software enables the user to select high-level quantum components that are needed for specific applications or tests. These high-level components are compiled by the software into a network configuration file that represents the physical configuration of lower-level equipment. The compiled network configuration is instantiated on the network by a second software package – the Control Plane software. The Control Plane software validates a network configuration before using NETCONF/YANG to configure and monitor the network's available equipment. The Control Plane software is responsible for verifying network performance, run-time monitoring of all network equipment, and coordination of the storage and transfer of archived data.

The interplay between the two software packages is illustrated in Figure 8.

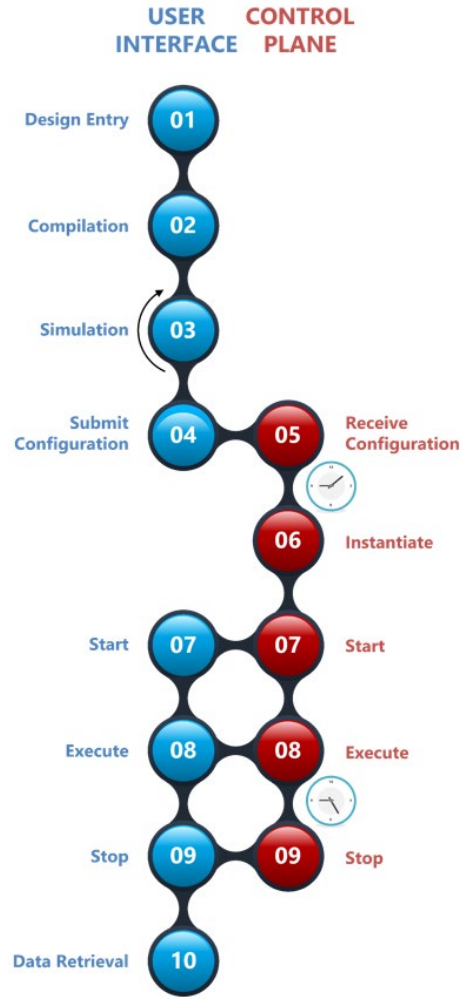

Fig. 8. **Software control of network:** The process of network design capture, compilation, instantiation, run-time data collection, and data retrieval are illustrated along with the software package (i.e., the User Design & Interface software and Control Plane software) responsible for each.

Once a network configuration is instantiated and active, data is generated. The data can be divided into three categories:

- Signal data
- Count Data
- Archive Data

Signal data is primitive electronic data (e.g., LVTTL pulses) available to the user through signal connectors (e.g., SMA connectors) on the QNIC interface. Signal data is real-time data associated with photon detection events and/or timing synchronization events. Signal data can be generated by both the network's equipment and the user's equipment.

Count data is serial data available to the user through a serial port on the QNIC interface. Count data includes the total number of single counts and correlated counts measured by the network's configured Timing & Control Modules. Count data can be queried and received through a user's equipment (e.g., a microcontroller) or computer via a serial port connection to the QNIC. Count data can be used to support control algorithms, feedback loops, iterative measurements, etc.

Archive data represents all other data collected from the network during an instantiation. Time-stamped photon arrival times, network equipment settings, monitored temperatures and other operational and environmental data is collected and stored in an archive data file. This file is stored and made available to the user, for a limited time, through a protected data portal (e.g., an authenticated web-based data portal).

## VI. USE CASES

A commercial quantum network can accelerate the development, adoption, and integration of quantum products. By providing common infrastructure for quantum solutions, quantum product developers are not required to build complete end-to-end solutions.

For example, Quantum Key Distribution (QKD) is a well-established method for securely distributing symmetric keys between users. Without a commercial quantum network, a provider of an entanglement based QKD solution (i.e., the BBM92 protocol) would need to develop an entangled photon source, automatic polarization controllers to compensate for fiber optic rotations, timing equipment and protocols to synchronize correlated events, and would need to deploy extensive routing and communication equipment to support distribution of quantum keys to multiple users.

With a commercial quantum network, a vendor can develop and sell a much simpler QKD product. By leveraging the common-use resources of a commercial quantum network (such as the Bohr-IV Quantum Network presented in this paper), the vendor need only provide a "QKD receiver" that would integrate with a connected customer's QNIC and measure the polarization of a received entangled photon (in two orthogonal bases). When used with a defined network configuration, this device would establish secure symmetric keys between the user's Quantum Node and an associated, protected Equipment Hub. Likewise, other users on the network could use simple "QKD receivers" to establish their own keys with the network's Equipment Hubs. Finally, the network's Equipment Hubs could use their own common-use resources to establish secure keys between each and every Equipment Hub.

Using secure key transfer techniques previously demonstrated with trusted relay QKD [25], any user of the network could establish QKD secured communication channels with any other user on the network.

It should be noted that the users would not need to have "QKD receivers" provided by the same vendor. Different vendors could provide various receivers (suitable for use with the BBM92 protocol) and all would be compatible across the commercial quantum network. Given the greatly reduced complexity of the "QKD receiver," vendors outside of the traditional quantum industry (e.g., in the more established telecom industry) would be able to develop and provide these products and solutions.

The QKD use case is just one example of how a commercial quantum network can accelerate the development, adoption, and integration of quantum products. Other near-term use cases are anticipated. A significant advantage of a commercial quantum network is that it provides an efficient platform for use case discovery. By creating a reconfigurable quantum infrastructure, the network can be used to support both technology development and early application deployment.

Additional near-term use cases anticipated with a commercial quantum network include:

- Quantum component testing and characterization for vendors [26]
- Protected, precision timing for time-sensitive networking and applications [27, 28]
- Distributed quantum sensing [29]
- Quantum Random Number Generators (QRNG) for post-quantum cryptography [30]

The requirement that a commercial quantum network utilize only commercially available devices means that long-term use cases will be enabled as new commercial components (e.g., quantum memories) reach maturity and are integrated into future networks. As network capabilities increase, future commercial quantum networks can be expected to enable use cases associated with distributed quantum computing [31,32] and long-range secure communications [33-35]. The progressive nature of the technology commercialization and adoption in future quantum networks is detailed in Figure 9.

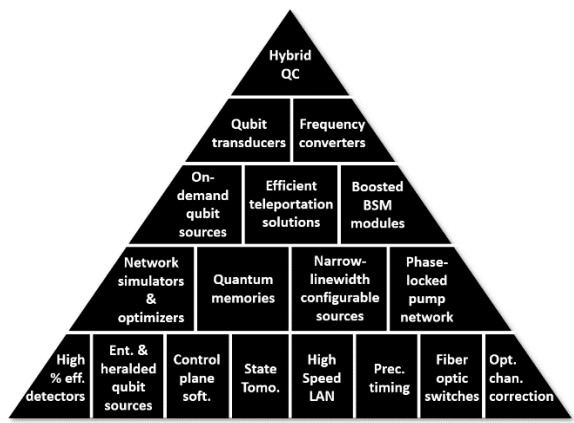

Fig. 9. **Emergence of new quantum network components and capabilities:** The bottom layer represents the state-of-the-art in commercial quantum components currently incorporated into the Bohr-IV Quantum Network. Each subsequent higher layer represents new commercial quantum products and capabilities accelerated through development, adoption, and integration into a commercial quantum network.

## VII. IMPLEMENTATION

At the writing of this paper, a first-generation commercial quantum network - based on the Bohr-IV Quantum Network architecture - is being constructed and deployed through an industry partnership. Future technical papers are anticipated that will detail the capabilities of the final deployed network as well as the performance of individual components. Likewise, resource sharing models and mechanisms that permit users to subscribe to and utilize the commercial quantum network are being developed.

## VIII. SUMMARY

In this paper we have described the architecture of a first-generation commercial quantum network based on commercially available devices. We have described the architectural topology of the network along with common-use resources provided by the network. We have detailed the connectivity between these devices as well as the software required to configure and monitor the network. We have provided both near-term and long-term use cases for the network and its potential impact on the development and commercialization of new quantum technologies. Finally, we have argued that the realization of these commercial quantum networks will accelerate the development, adoption, and integration of commercial quantum products.